# Quantum many-body principles of localized-state ensemble luminescence

Xinye Fan and Shijie Xu*

Photonics-X Laboratory, Department of Optical Science and Engineering, College of Future Information Technology, Fudan University, 2005 Songhu Road, Shanghai 200438, China
*Corresponding author. Email: xusj@fudan.edu.cn

## Abstract

Localized electron states induced by various disorders, including defects and impurities, usually exist in solids. Their electrical properties have been extensively investigated and well documented, while their optical properties such as localized-state ensemble (LSE) luminescence remain poorly understood. In particular, a microscopic quantum many-body (MB) theory has not yet been established for LSE luminescence so far. In this Letter, we attempt to fill this void via developing a quantum MB luminescence theory taking into account both electron-phonon (e-p) and electron-electron (e-e) interactions. Abnormal thermal behaviors such as redshift and subsequent blueshift of peak position, narrowing and succeeding broadening of linewidth, decline in intensity, and variation in lifetime can be quantitatively interpreted. Within the framework of the MB-LSE theory, moreover, Varshni's empirical formula for bandgap temperature dependence and Huang-Rhys factor for e-p coupling, and other key formulas are further derived and discussed.

## Introduction

Localized-state ensemble (LSE) luminescence is a ubiquitous physical phenomenon usually observed in luminescent solids with certain disorders [1]. For example, anomalous temperature-dependent luminescence behaviors of localized states were discovered as early as the 1960s [2]. They include an S-shaped shift in luminescence peak position and a non-monotonic variation in full width at half maximum (FWHM), etc. [3–6]. On one hand, the existence of localized states may be very beneficial to the performance improvement of LEDs, lasers, and other optoelectronic devices [7–11]. On the other hand, the existence of localized states also brings a challenge to the microscopic understanding and quantitative interpretation of LSE luminescence, even though a quantitative model based on the rate equation had been developed by one of the present authors and his team members [12, 13]. It is well recognized that LSE luminescence in solids is an important photophysical phenomenon involving extremely complicated many-body (MB) interactions. Unfortunately, to the best of our knowledge, a quantum mechanism-based MB microscopic theory has not yet been established for LSE luminescence in solids. Fortunately, the rapid development of quantum MB theory in recent years has brought new hope to achieve a possible solution to this long-term outstanding problem [14].

It has been well known that emissive optical transitions of localized carriers are inextricably linked with their transport processes [15, 16]. In this research, we would like to develop a microscopic quantum theory for LSE luminescence. First, an effective Hamiltonian was proposed for an LSE system with Anderson localized states and MB interactions including electron-electron (e-e) and electron-phonon (e-p) interactions [18]. Then, the time-resolved spontaneous emission probability of the localized carriers in LSE system was derived by finding the density matrix of the system.

Furthermore, steady-state luminescence of LSE system was yielded. Experimental data were collected for five material systems: InGaN quantum wells [4], ZnSeTeS quantum dots [8], copper-iodide hybrids [9], GaAs nanowires [19], and PbS quantum dots [20]. The temperature-dependent peak position, FWHM, integrated intensity and lifetime of LSE luminescence were then quantitatively interpreted. Several puzzling mechanisms, including the roles of e-p coupling and e-e interactions in the temperature-dependent LSE luminescence, were elucidated for the first time. Meanwhile, the Varshni's empirical formula for band gap shrinking and the Huang-Rhys factor for e-p coupling, and other key formulas describing the thermal behavior of solid-state luminescence were derived from the developed MB-LSE luminescence theory. Therefore, the establishment of such an MB-LSE microscopic theory may produce a novel and general understanding of the complicated LSE luminescence in solids and bridges different classic models of luminescence thermodynamics.

## Theoretical model

In solids, an LSE system considered here may comprise many Anderson localized states and extended states. In such a system, quasi-particles such as electrons, holes, excitons, etc. can move by themselves, recombine radiatively and non-radiatively, and can be scattered by phonons or can be captured and released by the localized states, as schematically illustrated in Fig. 1. In path $N$, a quasi-particle can get in and out of the LSE region. Although this particle may experience a complicated transport procedure in path $N$, it did not produce light emission. In path $N+1$, however, light emission, i.e., a photon generation, takes place.

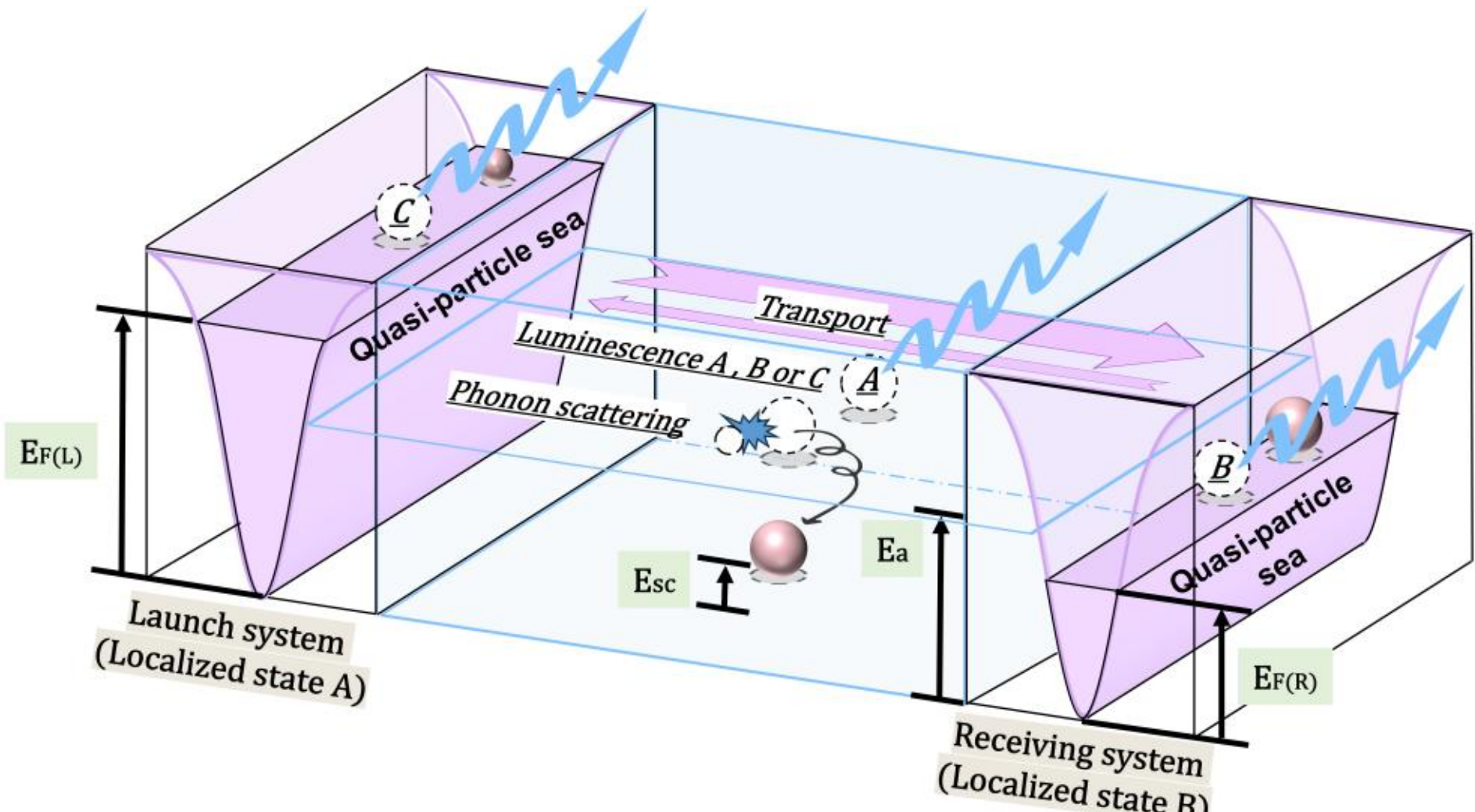

Fig. 1 A schematic picture illustrating transport and luminescence of quasi-particles from the launch source to the receiving terminal via LSE system in the phonon field. Launching quasi-particles may have a certain possibility of radiative recombination to produce luminescence in the LSE region. During the transport and luminescence processes, the quasi-particles may be scattered by phonons. The chemical potential energies of the launch system and receiving system are $E_{F(L)}$ and $E_{F(R)}$, respectively, while the tunneling level of the LSE is $E_a$.

As schematically shown in Fig. 1, three processes that the quasi-particles launched from the launch system ($l$) may experience include: (I) Directly arriving at the receiving system ($r$). (II) Scattered by optical phonons ($ph(O)$) and acoustic phonons ($ph(A)$) to the levels $E_{sc(O)}$ and $E_{sc(A)}$, respectively, followed by radiative or non-radiative recombination, and/or continuously transport in LSE region. (III) Radiative recombination ($p$) in the LSE region. Disregarding quasi-particle spin and assuming that only one quasi-particle is allowed to pass through an energy level channel. Meanwhile, the energy levels $E_{sc(O)}$ and $E_{sc(A)}$ for phonon scatterings are assumed to be far away from the tunneling energy level $E_a$ of LSE system [18]. It should be noted that radiative recombination and phonon scattering of quasi-particles can occur at any point within the central LSE system. Of course, direct transport between the launch and receiving ends can take place. Under such circumstances, an effective Hamiltonian describing the transport, phonon scatterings, particle-particle interactions, and luminescence may be written as:

$$\widehat{\mathcal{H}} = \widehat{\mathcal{H}}_0 + \widehat{\mathcal{H}}_{e-ph} + \widehat{\mathcal{H}}_{e-e} + \widehat{\mathcal{H}}_{e-h}, \tag{1}$$

where $\widehat{\mathcal{H}}_0 = \sum_l E_a \hat{c}_l^\dagger \hat{c}_l + E_a \hat{c}_s^\dagger c_s + E_p \hat{c}_p^\dagger \hat{c}_p + E_{sc(O)} \hat{c}_{sc(O)}^\dagger \hat{c}_{sc(O)} + E_{sc(A)} \hat{c}_{sc(A)}^\dagger \hat{c}_{sc(A)} + \sum_r E_a \hat{c}_r^\dagger \hat{c}_r + \sum_h E_h \hat{h}^\dagger \hat{h} + \sum_{ph(O)} + E_{ph(O)}^{Ph} \hat{a}_{ph(O)}^\dagger \hat{a}_{ph(O)} + \sum_{ph(A)} E_{ph(A)}^{Ph} \hat{a}_{ph(A)}^\dagger \hat{a}_{ph(O)}$, $\widehat{\mathcal{H}}_{e-ph} = \sum_{ph} \Omega_{ph(O)}^{Ph} (\hat{c}_{sc(O)}^\dagger \hat{c}_s \hat{a}_{ph(O)}^\dagger + \hat{c}_s^\dagger \hat{c}_{sc(O)} \hat{a}_{ph(O)}) + \sum_{ph} \Omega_{ph(A)}^{Ph} (\hat{c}_{sc(A)}^\dagger \hat{c}_s \hat{a}_{ph(A)}^\dagger + \hat{c}_s^\dagger \hat{c}_{sc(A)} \hat{a}_{ph(A)})$, $\widehat{\mathcal{H}}_{e-e} = \sum_l \Omega_l \left( \hat{c}_l^\dagger \hat{c}_s + \hat{c}_l \hat{c}_s^\dagger \right) + \sum_r \Omega_r \left( \hat{c}_r^\dagger \hat{c}_s + \hat{c}_r \hat{c}_s^\dagger \right)$, and $\widehat{\mathcal{H}}_{e-h} = -\sum_h \Omega_{e-h} \hat{c}_p^\dagger \hat{h}^\dagger \hat{h} \hat{c}_p$. Detailed definitions of various quantities in the sub-Hamiltonians can be found in the Supplementary Information (SI) S1. Referring to the outstanding work by Gurvitz and Prager [21], two interrelated differential equations in terms of density matrices about the probability of a particle being undetected ($\sigma_{aa}$) and detected ($\sigma_{bb}$) in the transport space can be eventually obtained:

$$\begin{cases} \frac{\partial \sigma_{aa}}{\partial t} = -\Gamma_L \sigma_{aa} + \Gamma_R \sigma_{bb} + \Gamma_P \sigma_{bb} + \Gamma_{Ph(O)} \sigma_{bb} + \Gamma_{Ph(A)} \sigma_{bb} \\ \frac{\partial \sigma_{bb}}{\partial t} = \Gamma_L \sigma_{aa} - \Gamma_R \sigma_{bb} - \Gamma_P \sigma_{bb} - \Gamma_{Ph(O)} \sigma_{bb} - \Gamma_{Ph(A)} \sigma_{bb} \end{cases}. \tag{2}$$

In Eq. (2), $\Gamma = 2\pi\rho|\Omega|^2$ gives average width of the level $E_a$ due to interparticle interactions, where $\Omega$ is the coupling strength, and $\rho$ denotes the effective occupation density of states (DOS) for the corresponding particle. Derivation of Eq. (2) can be found in the SI document. By solving Eq. (2) with the initial conditions of $\sigma_{aa}(0) = 1$ and $\sigma_{bb}(0) = 0$, one can yield:

$$\sigma_{aa}(\mu, T, t) = \frac{X - \Gamma_L}{X} + \frac{\Gamma_L}{X} e^{-tX}, \tag{3}$$

$$\sigma_{bb}(\mu, T, t) = \frac{\Gamma_L}{X} - \frac{\Gamma_L}{X} e^{-tX},$$

where $X = \Gamma_L + \Gamma_R + \Gamma_P + \Gamma_{ph(O)} + \Gamma_{ph(A)}$.

It should be noted that for any quasi-particles, if they are successfully transported from the launch system to the receiving end, they would be unable to emit photons, no matter what complicated processes they experienced. For a given quasi-particle, its $N + 1$ path consists of the LSE region, although it may have undergone multiple transport events between lattice points before entering the LSE zone. Now let us consider a quasi-particle undergoing transport ($tr$), entering the LSE space, and subsequently experiencing phonon scattering ($sc$), radiative recombination ($p$), and re-excitation ($re$). If the total duration of this process coincides precisely with the time required for the LSE luminescence intensity to decay to its $e^{-1}$, then occurring probability (natural units $\hbar = 1$) of this event may be written as:

$$\boldsymbol{\mathcal{P}}(\boldsymbol{\mu}, \boldsymbol{T}, \boldsymbol{t_{tr}}, \boldsymbol{t_{sc}}, \boldsymbol{t_p}, \boldsymbol{t_{re}}) \propto \sigma_{bb(tr)}(\mu, T, t_{tr})^{Tr} \sigma_{aa(sc)}(\mu, T, t_{sc})^{Sc} \sigma_{aa(p)}(\mu, T, t_p)^{P} \sigma_{aa(re-e)}(\mu, T, t_{re})^{Re} \propto e^{-\mathcal{X} M t_e}, \quad (4)$$

where $Tr$, $Sc$, $P$, $Re$, and $t_{tr}$, $t_{sc}$, $t_p$, $t_{re}$ respectively represent the number of occurrences, and the time at which their respective physical processes take place, and $\mathcal{X} = \frac{Sct_{sc}+Ret_{re}}{Trt_{tr}+Sct_{sc}+Pt_p+Ret_{re}}\Gamma_L + \frac{Ret_{re}}{Trt_{tr}+Sct_{sc}+Pt_p+Ret_{re}}\Gamma_R + \frac{CTrt_{tr}}{Trt_{tr}+Sct_{sc}+Pt_p+Ret_{re}} + \frac{Pt_p+Ret_{re}}{Trt_{tr}+Sct_{sc}+Pt_p+Ret_{re}}\Gamma_P + \frac{Sct_{sc}+Ret_{re}}{Trt_{tr}+Sct_{sc}+Pt_p+Ret_{re}}\Gamma_{ph(O)} + \frac{Sct_{sc}+Ret_{re}}{Trt_{tr}+Sct_{sc}+Pt_p+Ret_{re}}\Gamma_{ph(A)} = \Gamma_L{}' + \Gamma_R{}' + \Gamma_P{}' + \Gamma_{ph(O)}{}' + \Gamma_{ph(A)}{}'$. Here $C$ is a constant related to the e-e interactions during successful transport, and $t_e = Trt_{tr} + Sct_{sc} + Pt_p + Ret_{re}$ is the total time. If there are $N_l$ localized states whose transport and recombination can occur within unit time and the DOS distribution of the localized-state ensemble is $\rho(\mu)$, then an approximate expression for time-resolved LSE luminescence intensity may be expressed as:

$$\boldsymbol{\mathcal{J}_{t-r}} \propto \wp N_l \rho(\mu) \boldsymbol{\mathcal{P}}(\boldsymbol{\mu}, \boldsymbol{T}, \boldsymbol{t}) = \wp N_l \rho(\mu) e^{-\mathcal{X} t}. \quad (5)$$

Here $\wp = (\frac{\Gamma_P{}'}{\Gamma_P{}'+\Gamma_{ph(O)}{}'+\Gamma_{ph(A)}{}'})^{Re}$ represents the radiative recombination probability. Furthermore, integrated intensity of steady-state LSE luminescence may be calculated with:

$$\boldsymbol{\mathcal{J}_{ss}} \propto \wp N_l \rho(\mu) \int_0^{\infty} e^{-\mathcal{X} t_e} d\, t_e = \bar{\wp} N_l \rho(\mu) / \mathcal{X} . \quad (6)$$

$\bar{\wp} = (\frac{\Gamma_P{}'}{\Gamma_P{}'+\Gamma_{ph(O)}{}'+\Gamma_{ph(A)}{}'})^{\overline{Re}}$ denotes the average radiative recombination probability of all quasi-particles in the LSE ensemble, which arises from re-excitation processes. Here, $\overline{Re}$ represents the average number of re-excitation events. In addition, Gibbs free energy loss (GEL) of quasi-particles via emission of phonons should be considered. Therefore, photon energy $E$ of radiative recombination may be corrected to:

$$\boldsymbol{E} = \mu - 2(\omega_{ph(O)}.\Gamma_{ph(O)}{}' + \omega_{ph(A)}.\Gamma_{ph(A)}{}')\bar{t}_{lsc}, \quad (7)$$

where $\bar{t}_{lsc}$ is the average time since the last scattering event occurred, and its value should be on the same order of magnitude as the luminescence lifetime. If the quasi-thermal equilibrium state of quasi-particles is neglected, an analytical expression for the peak position can be derived (by referring to SI S2).

### Theoretical photoluminescence (PL) spectra and comparison with experimental spectra

To test applicability of the above-derived microscopic theory for LSE luminescence, we would like to compare theoretical PL spectra with available experimental spectra in literature. Fig. 2a shows the PL spectral data (hollow circles) of the ZnSeTeS quantum dots by Wu et al. [8]. Solid lines represent the theoretical spectra calculated with the new theory (denoted as MB-LSE model here and after). Agreement between theory and experiment is satisfactory for such a complicated solid system. It is also found that the PL line profiles become relatively symmetric with increasing temperature. This may reflect the fact of thermal redistribution of the quasi-particles within localized states [13]. Temperature-dependent FWHM, peak positions, integrated intensities, and decay times of both theoretical (solid lines) and experimental data (hollow circles) from Wu et al. (ZnSeTeS quantum dots) [8], Cho et al. (InGaN quantum wells) [4], Zhu et al. (copper-iodide hybrids) [9], and H. Akiyama et al. (GaAs nanowires) [19], respectively, are shown in Figs. 2b to 2e. S-shaped temperature dependence of the peak positions and distinct nonmonotonic variation of the FWHM and PL decay times can be well reproduced by the MB-LSE model. Figs. 2f and 2g depict calculated steady state and transient PL spectra for copper-iodide hybrids at 298 K [9]. In addition, taking a typical strongly disordered InGaN quantum wells as an example, the roles of e-p and e-e interactions in the temperature-dependent PL behaviors will be discussed as follows.

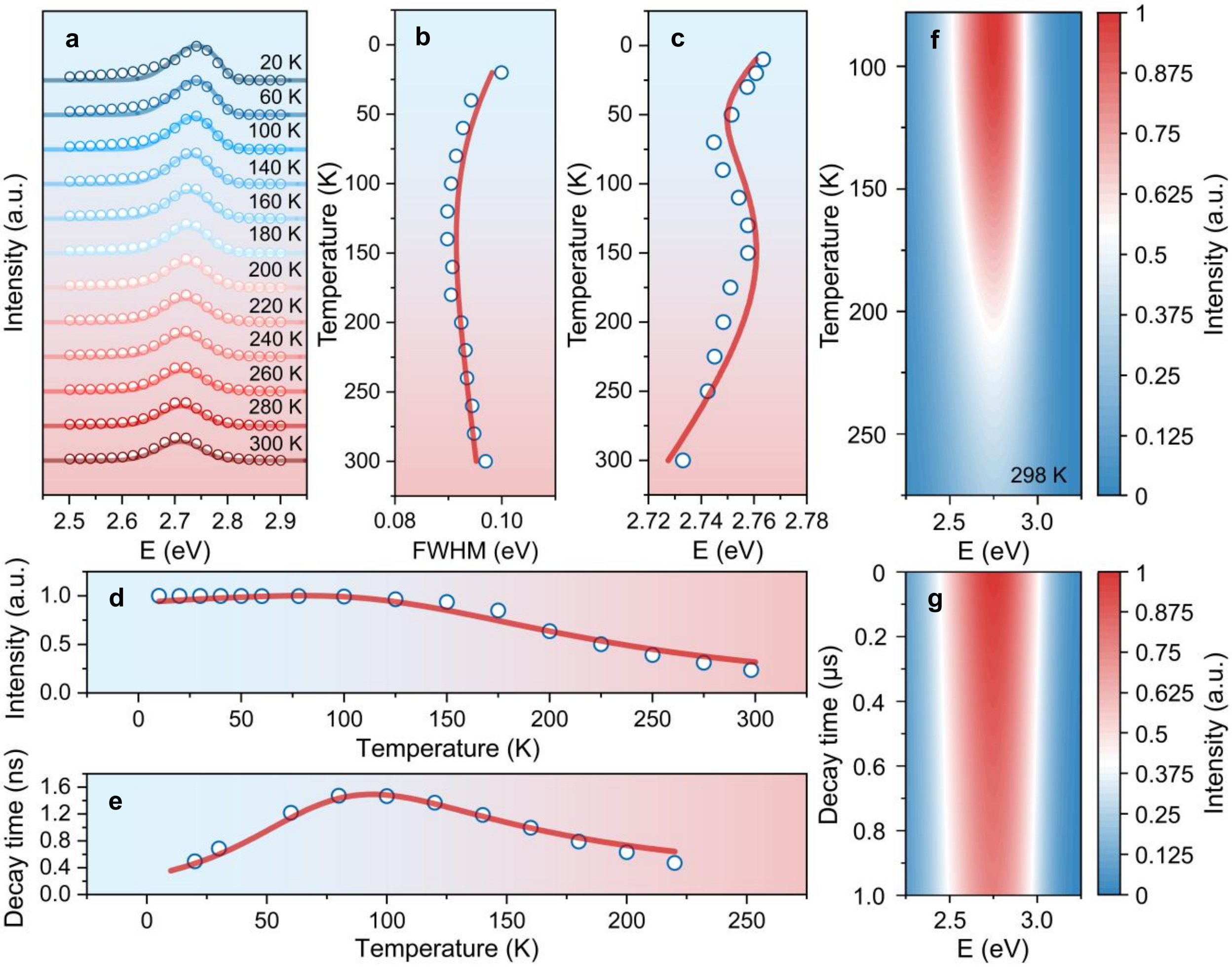

Fig. 2 Theoretical calculation and comparison with experimental spectra. (a) Theoretical PL spectra (solid lines) with the newly developed MB-LSE model and the experimental data (blue circles) of the ZnSeTeS quantum dots by Wu et al. [8] for different temperatures. (b) to (e) Temperature-dependent FWHM, peak positions, integrated intensities, and decay times (from top to bottom). Hollow circles represent the experimental data by Wu et al. (ZnSeTeS quantum dots) [8], Cho et al. (InGaN quantum wells) [4], Zhu et al. (copper-iodide hybrids) [9], and H. Akiyama et al. (GaAs nanowires) [19], respectively, and the solid lines stand for theoretical results. (f) and (g) Calculated copper-iodide hybrids' steady state and transient PL spectra at 298 K with the newly developed MB theory. For a complete calculation of the respective luminescence properties and the relevant calculation parameters, refer to SI S3.

First, the GEL effect was theoretically examined. If the interactions between quasi-particles and phonons were independent on their free energy, only temperature dependence of the peak position in the range of medium-high temperatures is affected, i.e., the peak position turns to blueshift at about 50 K and tends to saturate at high temperatures, as shown in dotted green line in Fig. 3a. Theoretical dependence without considering the GEL effect deviates significantly from the experimental data (hollow circles) by Cho et al. in the temperature range of 50-300 K. Relevant calculations are also performed for other materials, such as GaAs-based materials [19], and the results are shown in SI S3. If the e-p interactions were not considered, as shown in dotted blue line in Fig. 3, for peak position and FWHM (inhomogeneous broadening), the e-p interactions promote blue shift and broadening, respectively, by influencing the evolution of the wave function. For integrated intensity and luminescence lifetime, both theoretical integrated intensity and lifetime rapidly increase with increasing temperature in the case without the e-p interactions, which are in contrast with the experimental dependence (hollow circles). In the low temperature range < 50 K, the experimental lifetime growth may be interpreted as the weaker e-p interactions. Additional calculations are provided in SI S4. Therefore, the e-p interactions play a significant role in determining thermal behaviors of solid LSE luminescence.

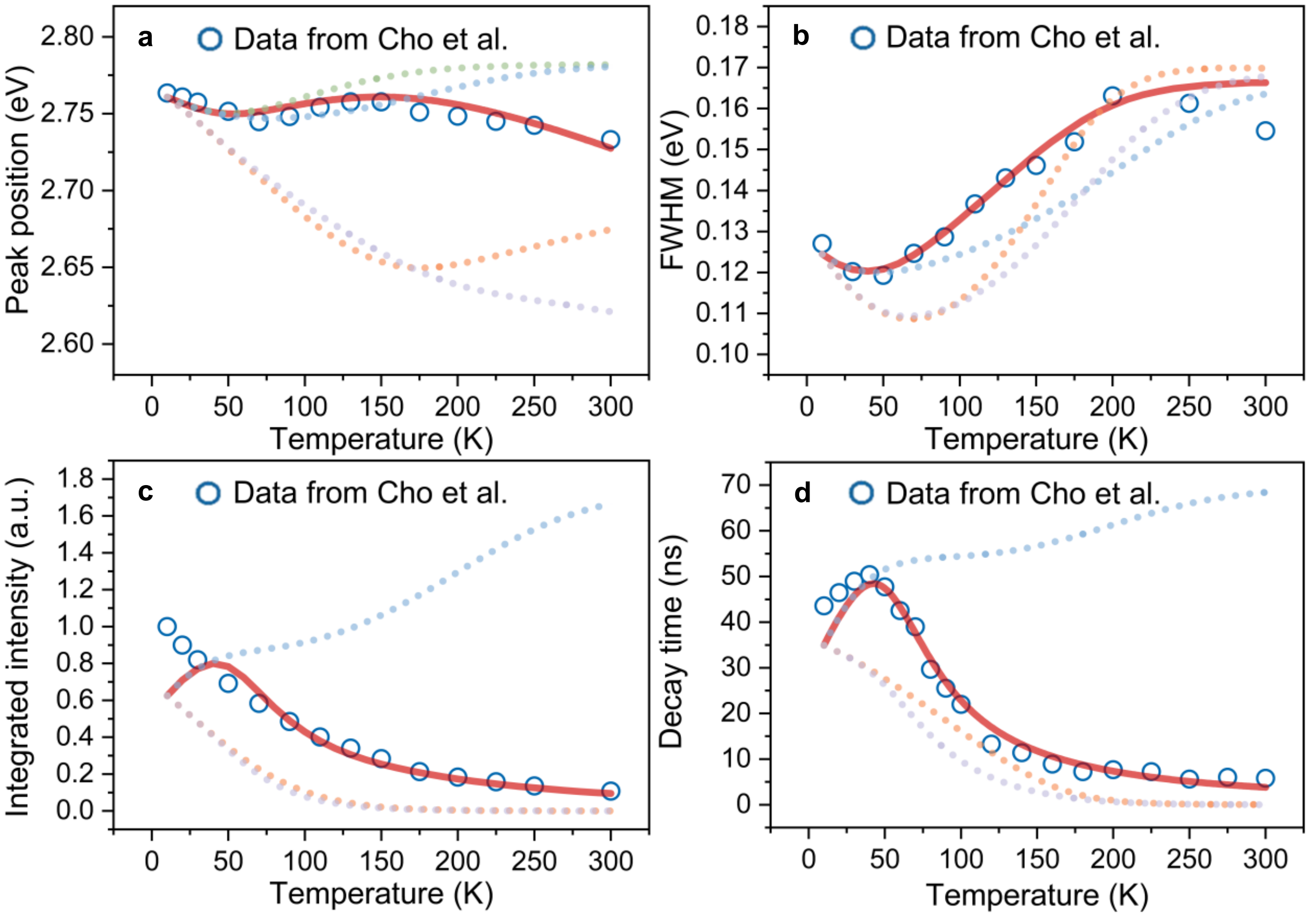

Fig. 3 Calculated temperature dependence of PL spectral parameters for several cases, i.e., with and without e-p and e-e interactions. (a) Peak position, (b) FWHM, (c) Integrated intensity, and (d) Decay time. Hollow circles represent the experimental data from Cho et al. [4].

Now it may be helpful for us to offer a possible microscopic explanation to the famous Varshni's empirical formula $E_{peak}(T) = E(0) - \gamma T^2/(\theta + T)$ for redshift temperature dependence of bandgaps of many materials [22]. By extremizing the Gaussian-type DOS distribution to a single energy level distribution and making a second-order Taylor expansion of the Eq. (7), we may obtain following approximate expressions: $\gamma \propto \left(\frac{Sct_{sc}+Ret_{re}}{Trt_{tr}+Sct_{sc}+Pt_p+Ret_{re}}\right) 4\pi\rho_{Ph|T=0}\left|\Omega_{\widetilde{ph}}\right|^2 \frac{\bar{t}_{lsc}}{\hbar} k_B = \boldsymbol{f} k_B$, where $\boldsymbol{f}$ is a dimensionless factor and $\theta \propto \frac{\hbar\omega_{\widetilde{Ph}}}{2k_B}$. Here, $\omega_{\widetilde{ph}} < w_o$, it is even close to the acoustic phonon frequency $w_a$ (i.e., Debye cut-off frequency) in certain material systems. Therefore, the e-p interactions are the major mechanism causing Varshni's redshift temperature dependence of the fundamental band gap in semiconducting and insulating materials. Regarding the e-p interactions, the well-known Huang-Rhys factor ($\boldsymbol{\mathcal{S}}$ factor) [23] shall be discussed because it is a pivotal parameter characterizing e-p coupling strength and hence determining some optical properties of solids [24]. On the premise that the Gibbs free energy loss associated with the quasi-particle is the sole contributor to the modification of lattice elasticity, we derive:

$$\boldsymbol{\mathcal{S}} = \boldsymbol{f}\bar{n}_{Ph}(\omega), \tag{8}$$

where $\bar{n}_{ph}(\omega)$ denotes the average number of phonons with angular frequency $\omega$. If we set $\boldsymbol{f} = 2\mathcal{S}_0$, then shift in peak position of LSE luminescence caused by GEL is: $\boldsymbol{\mathcal{S}}\hbar\omega_{\widetilde{Ph}} = 2\mathcal{S}_0\hbar\omega_{\widetilde{Ph}}\bar{n}_{ph}$, which is well consistent with the e-p coupling term in O'Donnell-Chen's empirical formula for redshift temperature dependence of bandgap [25]. The Huang-Rhys $\boldsymbol{\mathcal{S}}$ factor is thus a comprehensive function of temperature, phonon energy, and phonon scattering time. In addition, when $\sigma^2 \to 0$, S-shaped temperature dependence of peak position of LSE luminescence can diminish and tend to a normal Varshni-type temperature dependence, as shown in Fig. 4a. In fact, from Eq. (6), a maximum redshift amount $\Delta E_{LTR}$ can be shown to be less than $\xi \cdot \frac{(1-g)\sigma^2}{k_B T_e}$. Here $\xi = \frac{\Gamma_{L0}'}{\Gamma_{L0}'+\Gamma_R'+\Gamma_P'+\Gamma_{ph(O)}'+\Gamma_{ph(A)}'}$, $\Gamma_{L0}' = \frac{Sct_{sc}+Ret_{re}}{Trt_{tr}+Sct_{sc}+Pt_p+Ret_{re}} 2\pi\rho_{L|T=0}(E_a)|\Omega_L(E_a)|^2$, and other parameters' definitions can be found in SI S1.

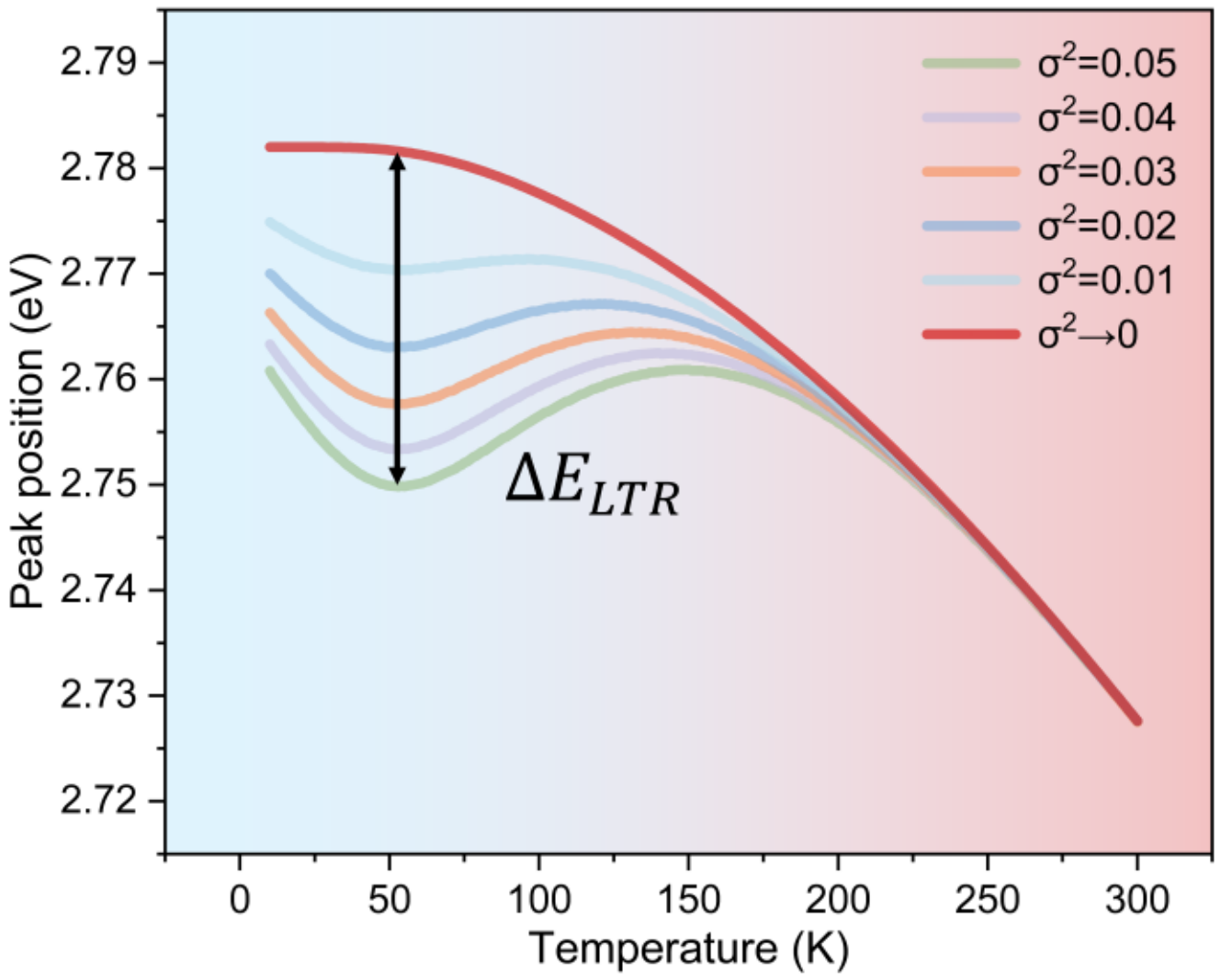

Fig. 4 With $\sigma^2$ decreasing from 0.05 to zero, S-shaped peak-position shift gradually transfers to a normal Varshni-type temperature dependence.

Besides the e-p coupling, the e-e interactions are another important factor affecting the properties of LSE luminescence. They can also induce variations in the width of the energy level via the thermal distribution of quasi-particles with disparate energies and the modulation of coupling strengths. It is revealed that variations in e-e coupling strength significantly influence the thermal properties of LSE luminescence. When variations in e-e coupling strength were not considered, temperature dependence of all four theoretical spectral parameters (dotted purple curves in Fig. 3), including peak position, FWHM, integrated intensity, and lifetime, substantially deviates from tendencies of the experimental data (hollow circles in Fig. 3). In addition, as detailed in SI S6 (Combined with PbS quantum dots data [20]), the Coulomb interactions between electrons also influence the injection and extraction of electrons in localized states. Finally, upon considering the combined effects of e-p and e-e interactions, temperature dependence of luminescence intensity of a two-level system can be approximated as: $\left[1+\zeta(T)e^{\frac{E_0-E_a}{k_B T_e}}\right]^{-1}$ and $\frac{{\Gamma_P'}^{\overline{Re}}}{(\Gamma_R'+\Gamma_P')^{\overline{Re}+1}}e^{-\frac{(\overline{Re}+1)\not{f}k_B}{(\Gamma_R'+\Gamma_P')\bar{t}_{lsc}\omega_{\widetilde{Ph}}}T}$ at low and high temperatures, respectively. Here $\zeta(T)=\frac{\Gamma_{L0}'}{\Gamma_R'+\Gamma_P'+\Gamma_{ph(O)}'+\Gamma_{ph(A)}'}$. For materials such as InGaN and InGaAs etc., there may exist $\Gamma_{ph(O)}'+\Gamma_{ph(A)}' \ll \Gamma_R'+\Gamma_P'$ at low temperatures and $\Gamma_R' \rightarrow 0$ at high temperatures, respectively. Then the LSE luminescence intensity can be approximated as the famous Arrhenius form at low temperatures, and it can also be approximated as the laser temperature characteristic formula with a distinct re-excitation process [26, 27].

## Conclusions

In summary, a microscopic many-body quantum theory is developed for LSE luminescence in solids. In this theory, an effective Hamiltonian considering e-e and e-p interactions was proposed and solved. The newly developed MB-LSE luminescence model is then applied to quantitatively interpret the available temperature-dependent steady-state and time-resolved PL spectra of the InGaN quantum wells, ZnSeTeS quantum dots, copper-iodide hybrids, GaAs nanowires, and PbS quantum dots. Temperature dependence of several key spectral parameters, including peak position, FWHM, integrated intensity, and lifetime, can be well elucidated. For the unusual blueshift of the luminescence peak frequently observed, theoretical analysis indicates that the variation in luminescence probability caused by e-p interactions plays a key role, whereas the redshift of peak position in high-temperature region usually originates from the loss in the Gibbs free energy. For the FWHM of LSE luminescence, phonon scattering causes broadening. Meanwhile, phonon scattering plays a major role in the integrated intensity attenuation and the lifetime decline with temperature, especially at high temperatures. It is also unraveled that e-e interactions, especially their variations with temperature, are an important factor influencing the LSE luminescence characteristics at low temperatures, for instance, resulting in red shift in peak position, narrower FWHM, and an increase in both integrated intensity and decay time. The MB-LSE theory developed in the present study can be used to derive expressions for several well-known formulas and parameters, such as Varshni's empirical formula and the $\boldsymbol{S}$ factor, etc.

## Data availability

All data are available in the main text or the supplementary materials.

## Acknowledgments

The authors wish to thank Dr. Debao Zhang, Dr. Ji Zhou, Dr. Wanggui Ye, Dr. Xuguang Cao, Mr. Sicheng Liu, Mr. Ke Yu, Mr. Zhenghao Guo, and Prof. Jiqiang Ning for their support and important contributions to the work. The work was financially supported by the National Natural Science Foundation of China (No. 12074324).

## Author contributions

Conceptualization: X. Y. F., S. J. X.; Methodology: X. Y. F., S. J. X.; Investigation: X. Y. F.; Visualization: X. Y. F., S. J. X.; Funding acquisition: S. J. X.; Project administration: S. J. X.; Supervision: S. J. X.; Writing - original draft: X. Y. F.; Writing - review & editing: S. J. X.

## Competing interests

Authors declare that they have no competing interests.